\documentclass[prl,reprint,twocolumn,superscriptaddress,amsmath,amssymb,amsfonts]{revtex4-1}

\usepackage{times}
\usepackage{graphicx,color}
\usepackage{hyperref}
\usepackage{braket}
\usepackage{multirow}
\usepackage{indentfirst}
\usepackage[FIGTOPCAP,raggedright,nooneline,bf,footnotesize]{subfigure}
\usepackage{paralist}
\usepackage[usenames,dvipsnames]{pstricks}
\usepackage{overpic}
\usepackage{array}
\usepackage{numprint}

\newcommand{\figref}[1]{Fig.~\ref{#1}}

\begin{document}


\title{Time and spectrum-resolving mutiphoton correlator for 300-900 nm}

\author{Kelsey D. Johnsen}
\affiliation{Institute for Quantum Computing and Department for Physics and Astronomy, University of Waterloo, 200 University Ave.~West, Waterloo, Ontario, CA
N2L 3G1}
\author{Piotr Kolenderski}
\email{kolenderski@fizyka.umk.pl}
\affiliation{Institute for Quantum Computing  and Department for Physics and Astronomy, University of Waterloo, 200 University Ave.~West, Waterloo, Ontario, CA
N2L 3G1}
\affiliation{Institute of Physics, Faculty of Physics, Astronomy and Informatics, Nicolaus Copernicus University, Grudziadzka 5, 87-100 Torun, Poland}

\author{Carmelo Scarcella} 
\affiliation{Dipartimento di Elettronica, Informazione e Bioingegneria, Politecnico di Milano, Piazza Leonardo da Vinci 32, I-20133 Milano, Italy}

\author{Marilyne Thibault}
\affiliation{Institute for Quantum Computing  and Department for Physics and Astronomy, University of Waterloo, 200 University Ave.~West, Waterloo, Ontario, CA
N2L 3G1}

\author{Alberto Tosi}
\affiliation{Dipartimento di Elettronica, Informazione e Bioingegneria, Politecnico di Milano, Piazza Leonardo da Vinci 32, I-20133 Milano, Italy}

\author{Thomas Jennewein} 
\affiliation{Institute for Quantum Computing and Department for Physics and Astronomy, University of Waterloo, 200 University Ave.~West, Waterloo, Ontario, CA
N2L 3G1}

\begin{abstract}


We demonstrate a single-photon sensitive spectrometer in the visible range, which allows us to perform time-resolved and multi-photon spectral correlation measurements. It is based on a monochromator composed of two gratings, collimation optics and an array of single photon avalanche diodes. The time resolution can reach $110$ ps and the spectral resolution is $2$ nm/pixel, limited by the design of the monochromator. This technique can easily be combined with commercial monochromators, and can be useful for joint spectrum measurements of two photons emitted in the process of parametric down conversion, as well as time-resolved spectrum measurements in optical coherence tomography or medical physics applications.

\end{abstract}

\maketitle

There are a great number of applications which benefit from the ability to characterize either low-intensity light or correlations between single photons. Some examples are single photon sources \cite{Lutz2013,Lutz2014,Mosley2007,Wasilewski2006}, time-resolved fluorescence spectroscopy, single molecule detection \cite{Ingargiola2013} and optical coherence tomography \cite{Dubois2004}.  These experiments often require knowledge of the spectral nature of individual photons, but it is difficult to get this information at the single photon level. Measurement techniques based on scanning detectors and the dispersive properties of materials have been employed to this end, but have certain disadvantages that need to be overcome. 

One can distinguish three main techniques used to characterize the spectral properties of single photons. 1) Typical commercial spectrometers, based on CCD cameras, are capable of performing measurements of low intensity light. However, this technique does not include timing information, which limits the applicability of this method when applied to single photons. Furthermore, it does not allow for multiple photon correlation measurements. 2) Alternatively, monochromators based on scanning single-photon detectors \cite{Poh2007} are used to collect spectral information. These devices are single-particle sensitive, but are incapable of measuring more than one piece of information at a time. As such, it is not possible to measure multi-photon spectra using this method. 3) A third method is time-multiplexed single-photon spectroscopy \cite{Avenhaus2009}, which is based on photon arrival timing measurements. The resolution of this method depends on the jitter of the single-photon detector, time stamping electronics resolution and dispersion of optical fibers. While this method is already hard to implement in the telecom band because of low dispersion, low detection efficiency and attenuation in single mode fibers, it is impossible to use this method in the visible range due to even higher attenuation.


\begin{figure}[ht]
\centering
\begin{tabular}{c c}
\multicolumn{2}{l}{\subfigure[Multi-photon spectrometer setup]{
{\begin{overpic}[width=0.9\columnwidth]{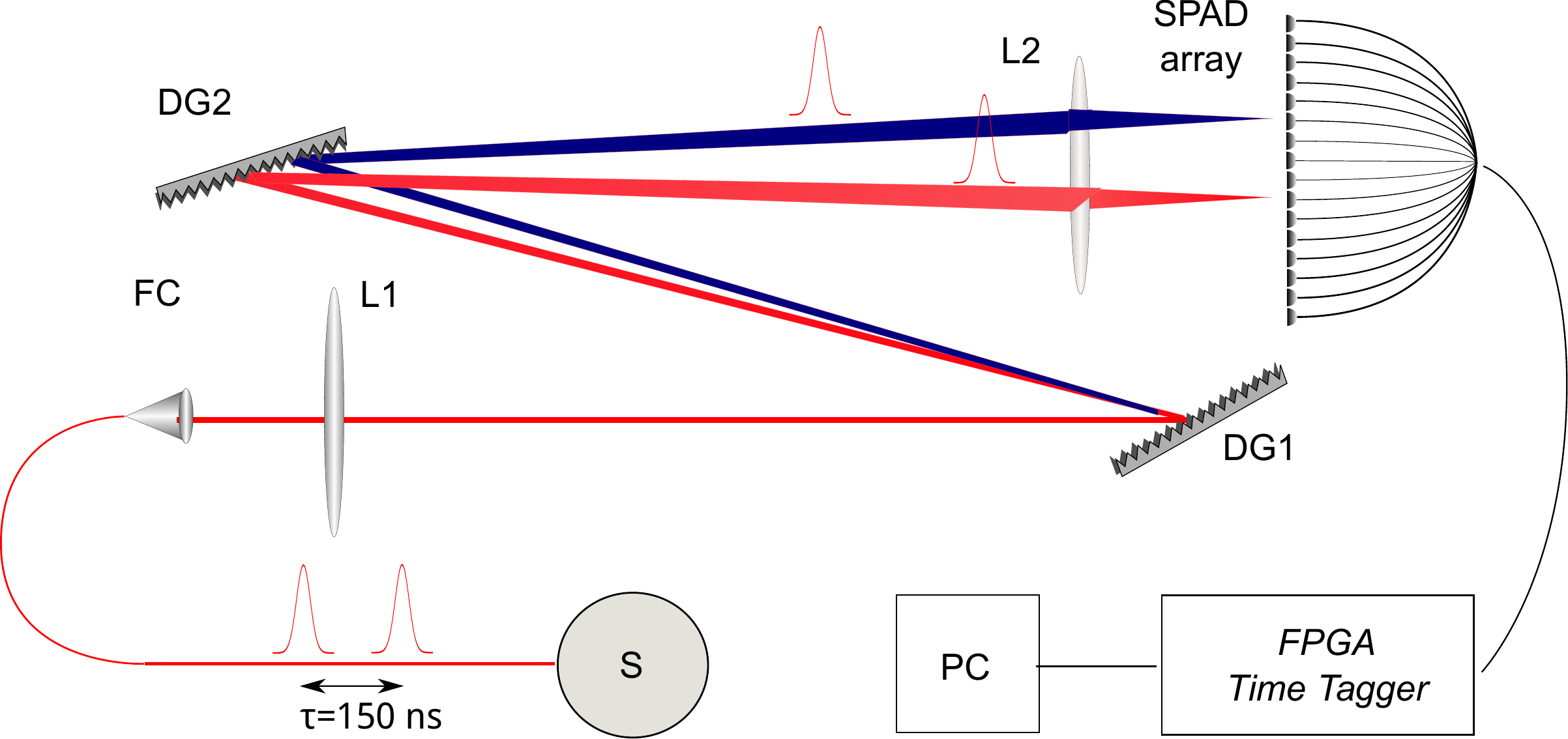}
\end{overpic}}
}}\\
\subfigure[]{\includegraphics[width=0.45\columnwidth]{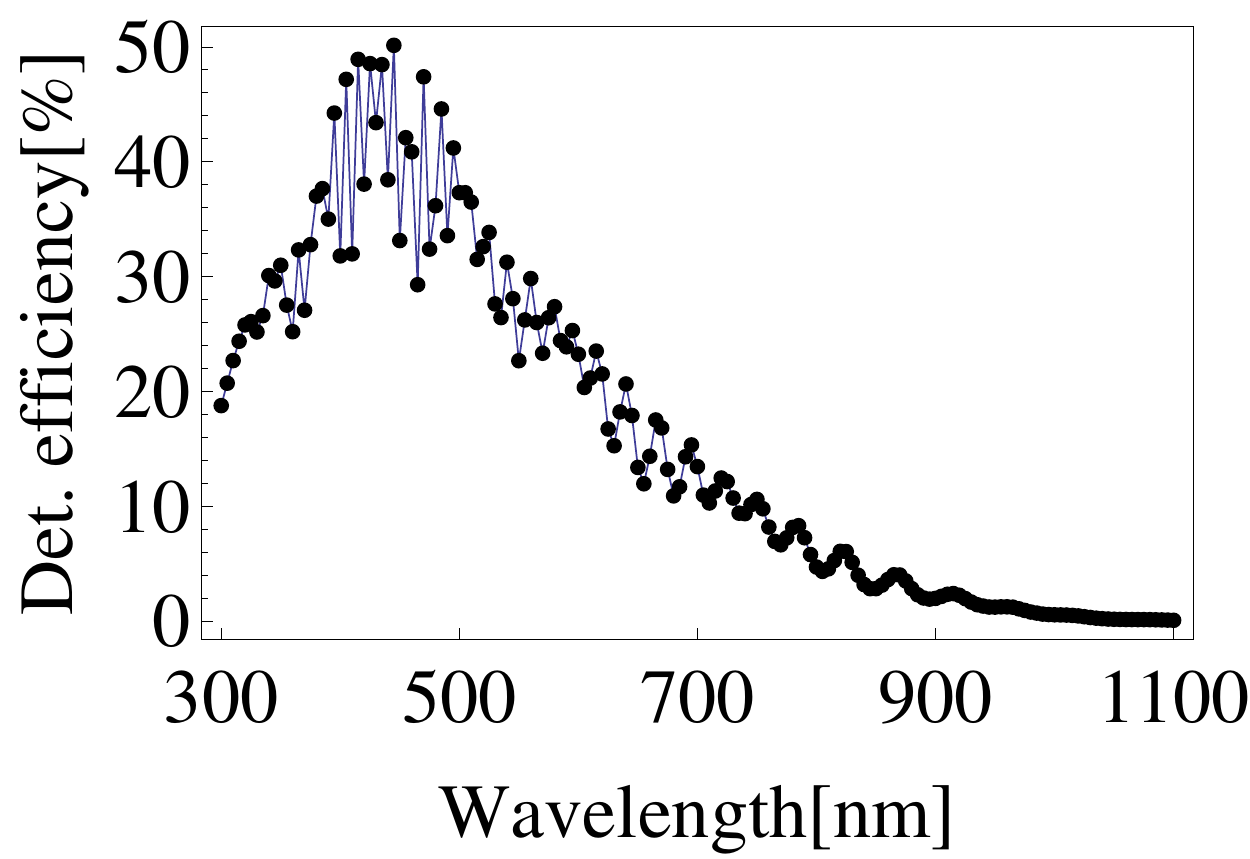} }&
\subfigure[Photo of the SPAD array]{\includegraphics[width=0.45\columnwidth,angle=180]{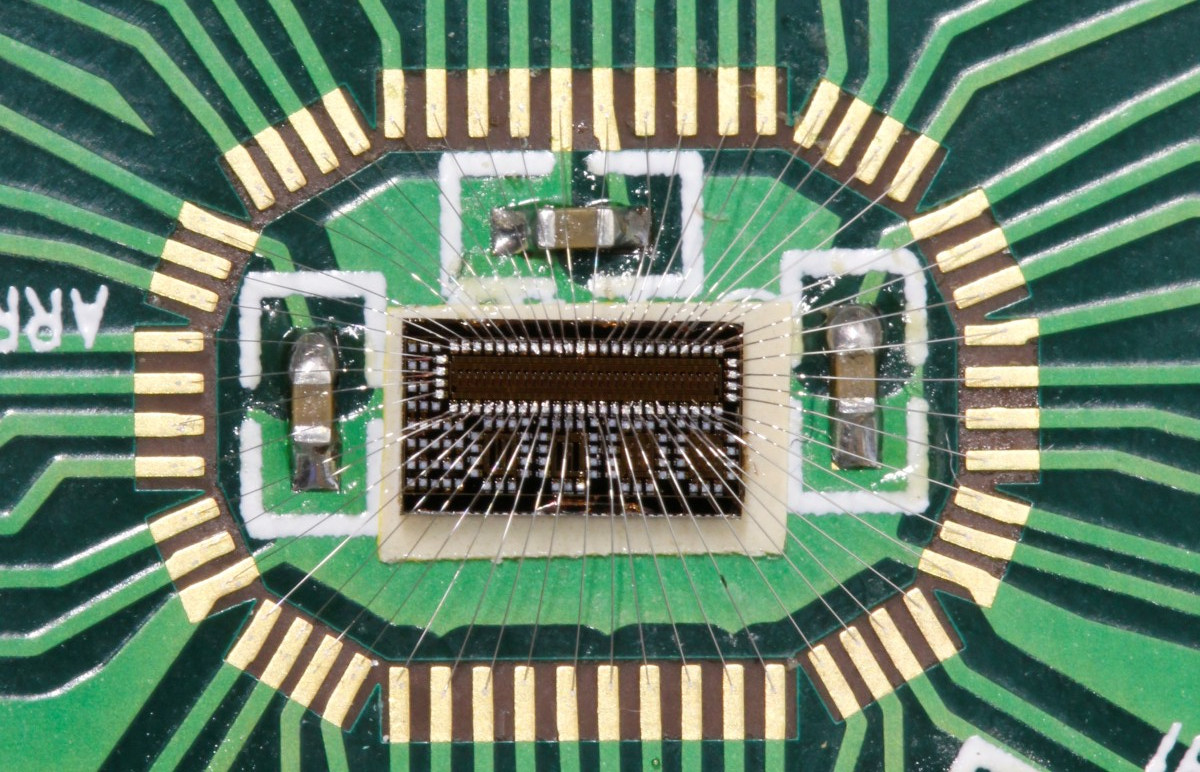}}
\end{tabular}
\caption{a) The multi-photon spectrometer experimental setup. Photons from an exemplary pair source S are delayed with respect to each other by $\tau = 150$ ns and sent to the spectrometer through single mode fiber using FC connector. Next, they travel through lens L1 (f$=35$ mm), which collimates the spatial mode; two diffraction gratings DG1 and DG2 ($1200/$mm at $750$ nm) and a lens L2 (f$=55$ mm), which focuses the mode on the SPAD array. Next, individual signals from the detectors are registered by FPGA time tagging unit. The time and channel number data is post-processed in a computer. b) The photon detection efficiency of the SPAD array. Note that the ripple of the photon detection efficiency curve is due to interference effects of the light in the passivation layer deposited above the detection chip \cite{Zappa2007,Scarcella2013}. c) The photo of the SPAD array detector.}
\label{fig:setup}
\end{figure}


Here we present the design and characterization of a multi-photon spectrometer for $300$-$900$ nm. Timing resolution allows us to measure multiple photon spectral correlations. The device is simple -- it consists of gratings and an array of 32 temporally- and spatially-resolved single-photon avalanche diodes (SPAD) \cite{Zappa2007,Scarcella2013}, shown in \figref{fig:setup} (a,c). The system of two gratings maps the wavelength of the incident photon onto the pixels of the array. 


There are two main applications for the device: measurements of single-photon spectra and spectral correlations between multiple photons. For the first, one needs to acquire the statistics of detection events for identical photons. This spectrometer measures the arrival time of photons, which can be used to evaluate the spectral correlation of two or more photons. For this application, the following steps must be followed: 1) if the goal is to measure the statistics of correlations between two photons, each photon should be delayed with respect to the previous photon by at least the hold-off time of the detector (the time after a detection during which the detector is off), 2) the location and time of detection are collected as a list of detector numbers and time stamps and 3) finally, the data is processed. This processing consists of looking through the time tags of detections and keeping the ones that arrive within an expected time range. Based on this data, the correlation can be computed as a histogram of pairs of detector numbers that clicked (in the case of two photons). 





The timing resolution of the device depends on both the SPAD array and time stamping electronics. The detector has a timing jitter of $110$ ps, and the FPGA electronic time tagging unit records the timing information with $156$ ps resolution. 

The spectral resolution of this implementation and the range of the multi-photon spectrometer is determined by the gratings. The SPAD array can detect photons in the range $300$-$900$ nm. For this demonstration, we set our grating system in the range of $790-830$ nm so that it is compatible with our single-photon sources. Due to the limited spectral resolution of the gratings, we only use $16$ pixels. This corresponds to a resolution of approximately $2$ nm/pixel. 

The overall photon detection efficiency is related to the quantum efficiency of the SPAD array, which is plotted in \figref{fig:setup} (b), and the transmission of the optical system. The quantum efficiency of the SPAD depends on the wavelength and reaches $50$ \% for $450$ nm. In our experiment, we use two single-photon sources (discussed in detail later in this paper), whose photons are centered around $810$ nm. The single-photon detection efficiency at this wavelength is around $5$ \%. The transmission of the grating system is approximately $50$\%. Thus, the overall photon detection efficiency of our spectrometer is around $2.5$ \%.  The probability of detecting $n$ photon coincidences is $(2.5\%)^n$. For a photon pair, this is approximately $0.06$\%.


In this setup, noise comes from two sources: dark counts and the afterpulsing effect. The dark count rate of the SPAD array is very low, approximately $100$ counts per second/pixel at room temperature. For the photon spectrum measurement, the dark count statistics can be used to subtract noise from the detection statistics. In the case of photon correlation measurements-- when multiple photon detection events are analyzed-- noise originating in dark counts is negligible. However, these correlation measurements can be significantly affected by the afterpulsing effect. This issue is one of the main problems requiring careful attention \cite{Zappa2007}. During the detection of a photon, many charge carriers flow through the SPAD. Some of these are captured by deep-level traps inside the depleted region of the detector, and are then released. If a charge carrier is released when the bias voltage is above the breakdown voltage, it could re-trigger an avalanche and cause a fake count, which would then be counted as a coincidence of two photons incident at the same detector. In order to reduce this correlated noise, the SPAD has to be off for a specific amount of time after each avalanche. This hold-off time is on the order of tens of ns for the SPAD array used in this spectrometer. By using a $150$ ns delay between the photons and setting a long hold-off time ($140$ ns), we decrease the afterpulsing probability per ns to as low as $10^{-4}$\%.

Cross-talk between neighbouring detectors is another problem when taking time-resolved correlation measurements. After a photon is detected, the SPAD pixel emits new photons which in turn can be detected by neighboring pixels. This effect is negligible and can be eliminated entirely by introducing a time delay between photons. Cross-talk events usually show up one after the other, since these events occur in a time interval of about $1.5$ ns after the avalanche ignition. Therefore, cross-talk does not impact our measurements, since we look for coincidences between events with a time-distance of $150$ ns.

The photon detections is limited predominantly by the hold-off time. The photon detection efficiency of the system is constant for low rates, and decreases for high count rates because the detector count rate saturates. With a hold-off time of $140$ ns, the detection rate of each pixel is linear with the impinging photon flux up to approximately $1$ MHz \cite{Scarcella2013,Zappa2007}. 

\begin{figure}[h]
\centering
\begin{tabular}{c c}
\subfigure[]{\includegraphics[width=0.48\columnwidth]{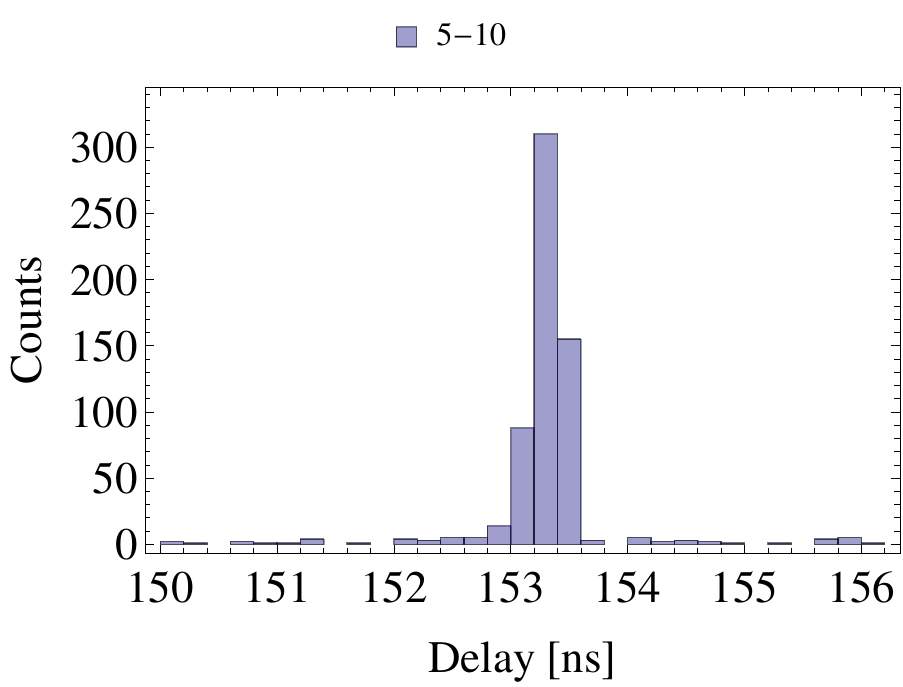}} &
\subfigure[]{\includegraphics[width=0.48\columnwidth]{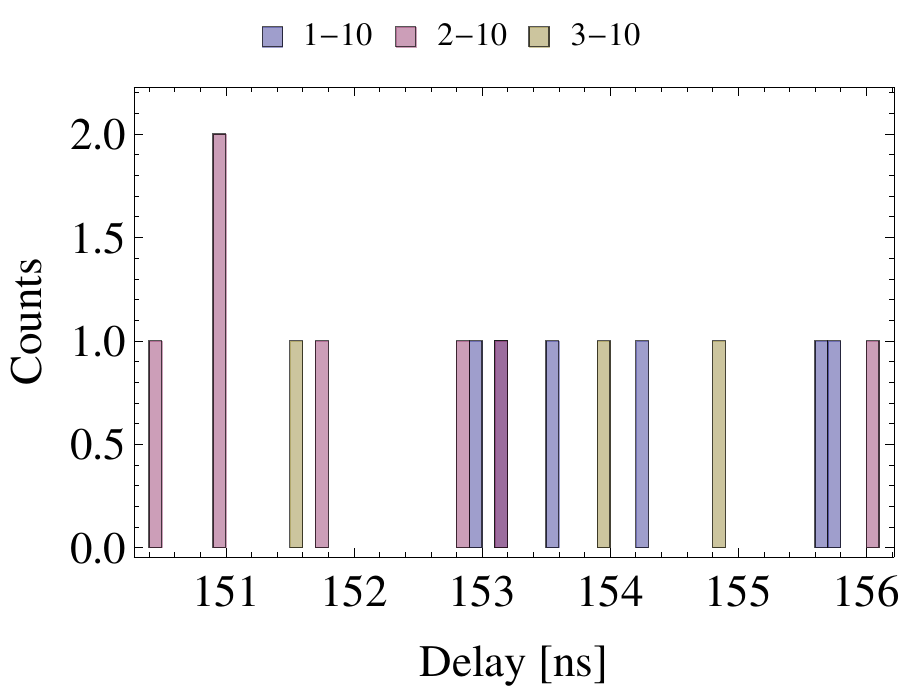}} \\
\subfigure[]{\includegraphics[width=0.48\columnwidth]{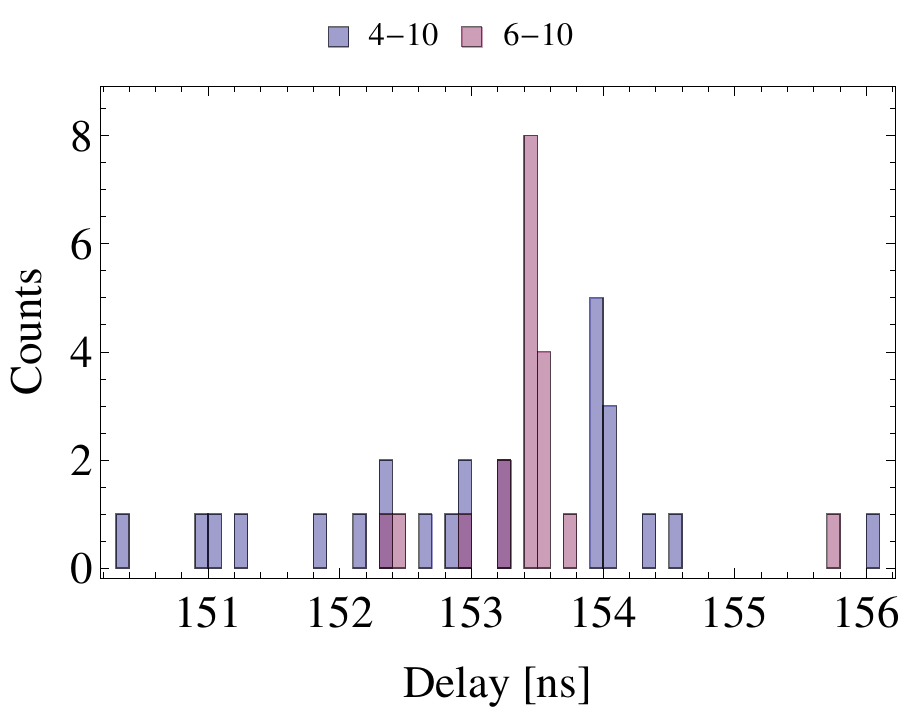}} &
\subfigure[SPDC pair spectral correlation] {\includegraphics[width=0.45\columnwidth]{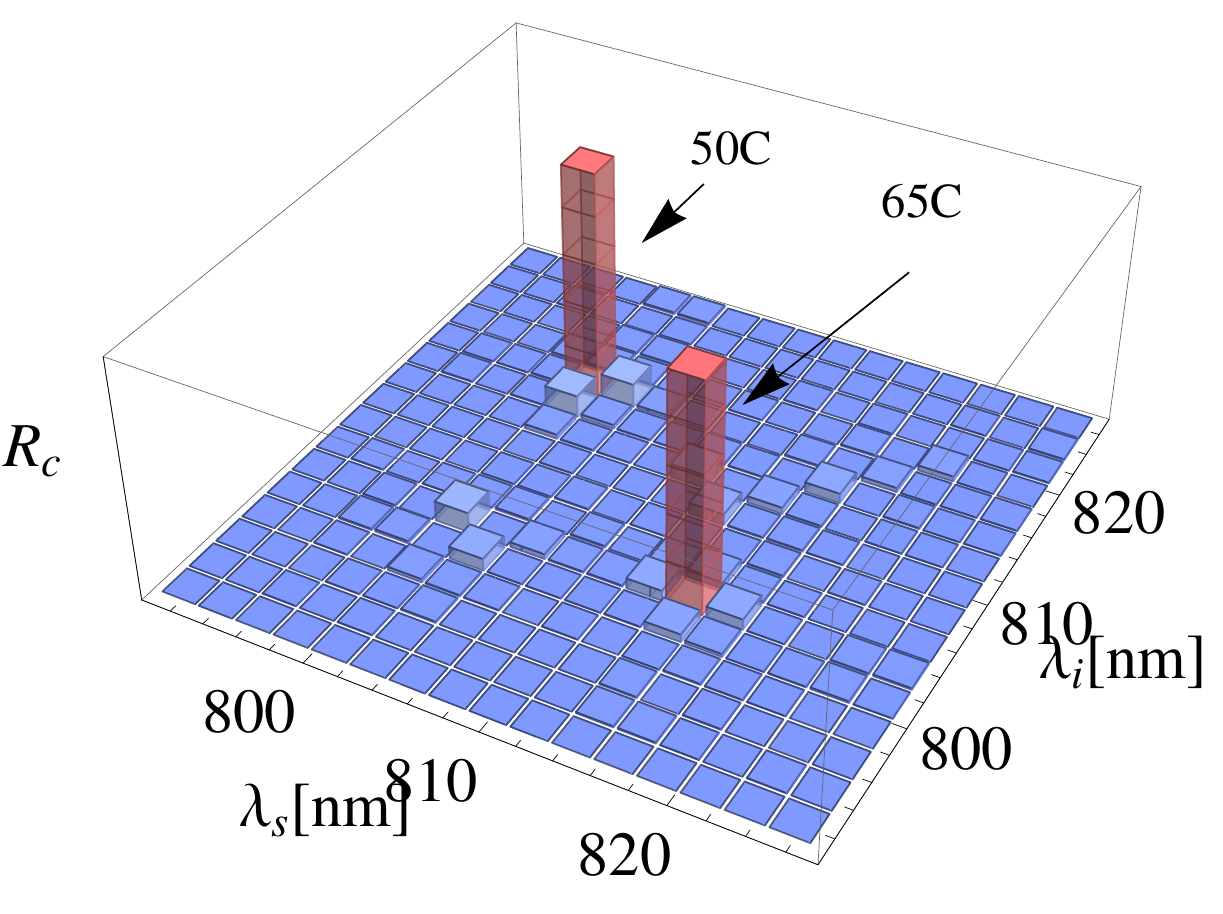}} 
\end{tabular}
\caption{Exemplary time histograms for SPDC source set at the temperature $50^o$C. The coincidences histogram for a channel pair a) 5-10 shows clearly visible peak consisting of around $600$ counts in a range of $500$ ps, b) only accidental coincidences totaling in 16 counts for channels 1-10, 2-10, 3-10 and c) small peaks for channels 4-10 and 6-10, which are related to finite resolution of the spectrometers.  d) Photon pair coincidences for SPDC source for two exemplary temperature settings $50^o$C and $65^o$C, corresponding to the first (second) photon's central wavelength, 802nm (814nm) and 813nm (803nm), respectively. } 
\label{fig:time:hist}
\end{figure}

We move on to the characterization of the muti-photon spectrometer using two single-photon sources emitting photons with known characteristics. We perform two tests using 1) frequency-correlated narrow-band single photon pairs emitted by a spontaneous parametric down-conversion (SPDC) source \cite{Fedrizzi2009} and 2) spectrally broadband coherent states emitted by an attenuated femtosecond laser (TiF50M, Atseva).

%

We use the SPDC source, which has two single mode fiber (SMF) outputs through which the photons propagate, to test the spectrometer's capability of measuring multi-photon correlations. We characterize the temperature tunability of our photon pair spectra with a commercial spectrometer by looking at each SMF output. The source produces frequency-degenerate photons at $809.6$ nm when the temperature is \mbox{$59^o$ C}. By tuning the temperature in the range $40^o$C$-70^o$C, we spectrally separate the photons by up to $15$ nm. The photons are initially coupled into two distinct SMFs. For testing purposes, we send one photon through $30$ m of SMF, which results in a delay of $150$ ns. The two photons are then combined by a polarizing beam splitter and coupled into SMF, which is then attached to the multi-photon spectrometer.

We acquire photon detection timing information for $60$ minutes at a given temperature. We repeat this measurement for two temperature settings $50^o$ C and $65^o$ C. This data is then post-processed to collect coincidences, and any photon pairs that arrive at the SPAD detector array with a delay of $150$ ns between each other are collected. The detections resulting from dark counts, afterpulsing or stray light are discarded. From this data, we plot a set of histograms, which convey information about the number of coincidences between detector pairs. These histograms are shown in \figref{fig:time:hist} (a-c). One sees only accidental coincidences for channel pairs 1-10, 2-10 and 3-10 as plotted in \figref{fig:time:hist}(a). However, there is a clearly visible peak for channels 5-10 in \figref{fig:time:hist}(b). The resolution of our spectrometer is finite, which results in few coincidences between channel pairs 4-10 and 6-10, which we depict in \figref{fig:time:hist}(c). Next, by looking at the time histograms, we compile a list of detector number pairs where the first of the pair corresponds to the photon which arrived first, and the second corresponds to the following photon. We plot coincidences as a histogram of the first photon channel versus the second photon channel. The SPDC source we use for the characterization is narrowband. This in combination with the resolution of our spectrometer results in single point to represent the pairs which occur most often. To show the capabilities of our spectrometer and for the clarity of presentation, we plot the results for two exemplary temperature settings $50^o$ C and $65^o$, see \figref{fig:time:hist}(d).

\begin{figure}
\centering
\begin{tabular}{c c}
\subfigure[puled laser spectrum] {\includegraphics[width=0.44\columnwidth]{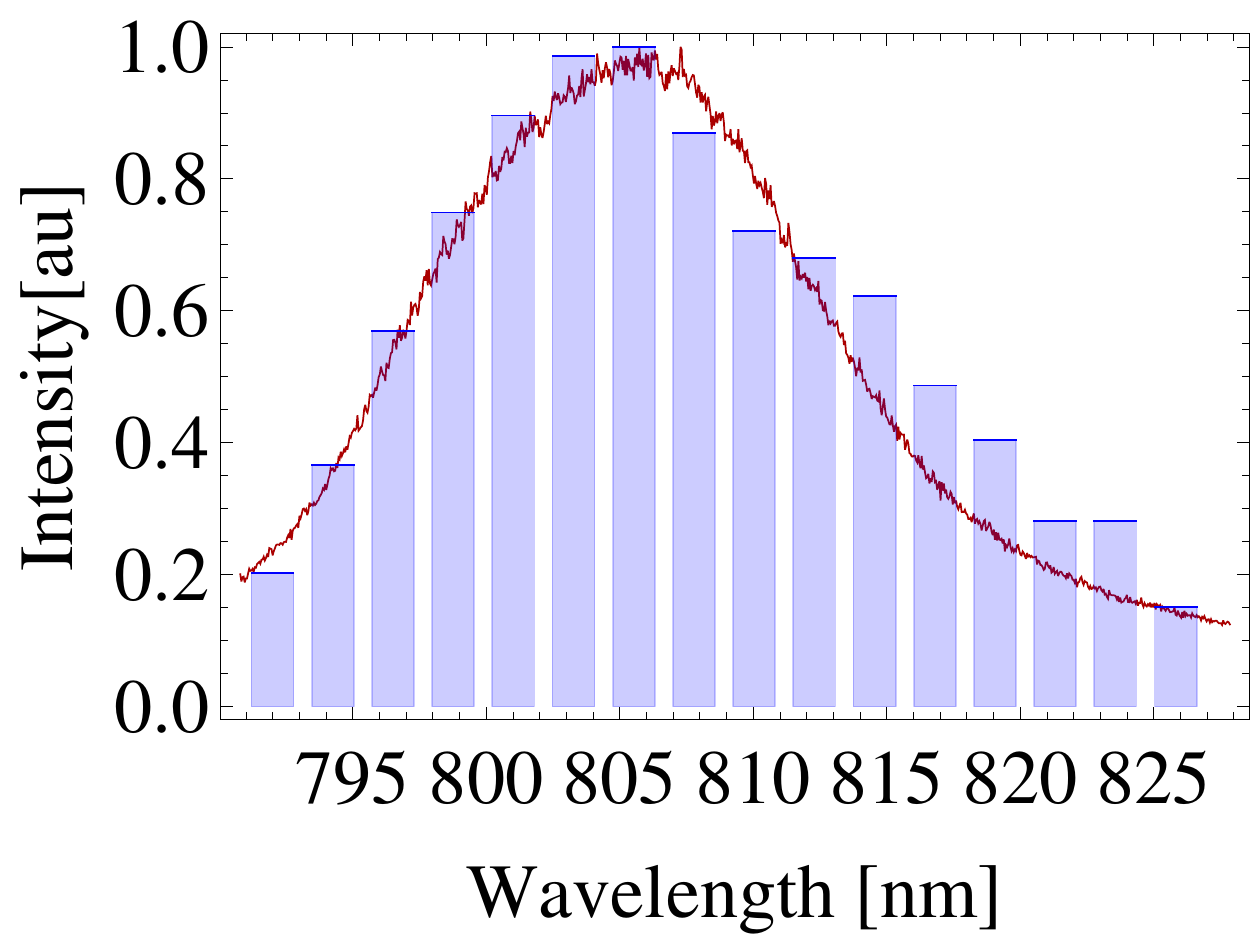}} &
\subfigure[two pulses correlation] {\includegraphics[width=0.45\columnwidth]{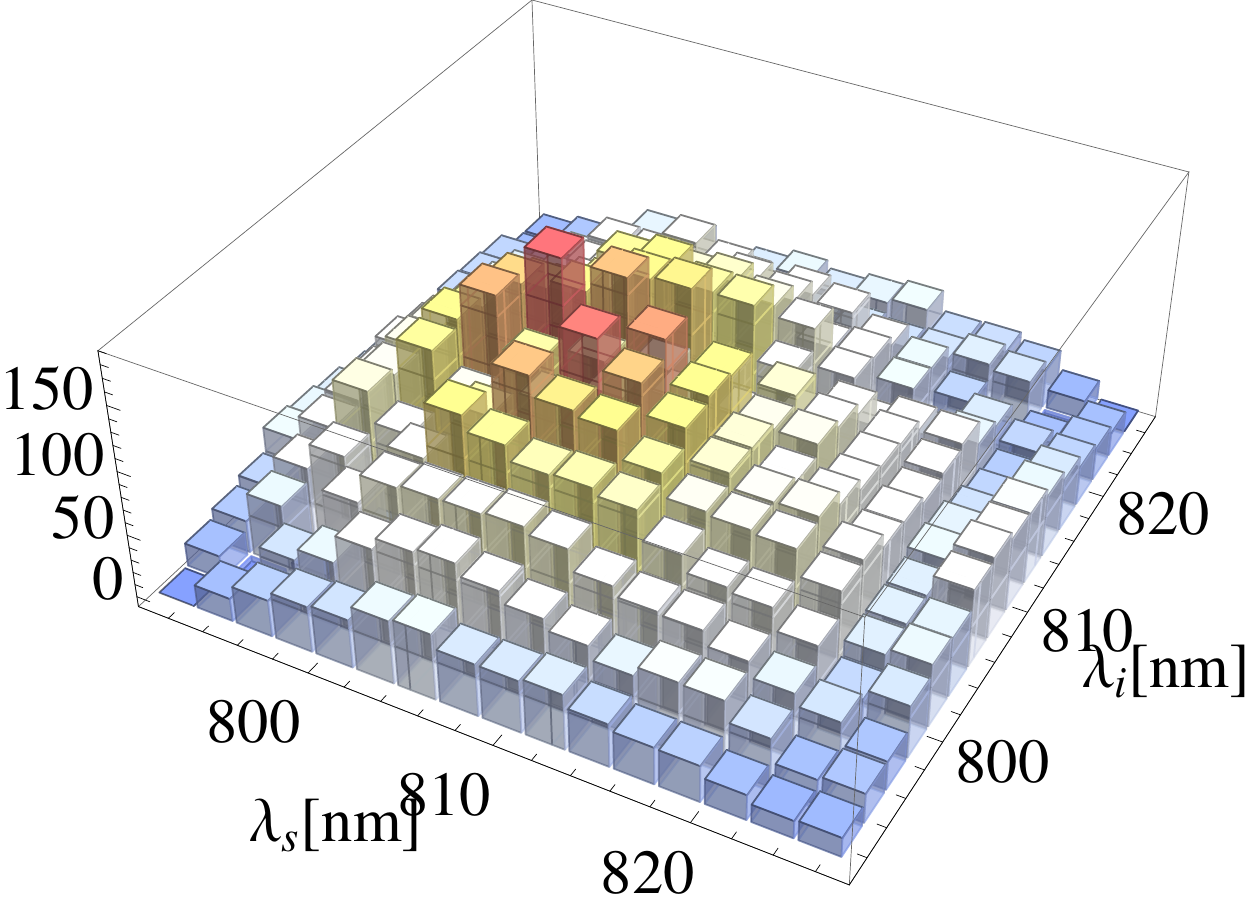}}
\end{tabular}
\caption{a) The spectrum of the coherent state generated by a Ti:Sapphire femtosecond laser attenuated to the single photon level. The bar plot shows the photon counts measured by the multi-photon spectrometer and the blue (online) line represents the spectrum measured by the commercial spectrometer. b) Spectrum correlation of two photons from coherent states. Here, the bar plot corresponds to the SPAD array counts and the red curve corresponds to the spectrum using a CCD-based spectrometer. } 
\label{fig:TIF}
\end{figure}

Next, we move on to the spectrally broadband pulsed source in order to test the accuracy of our intensity measurements. We use $67$ fs pulses from a Ti:Sapphire laser, attenuated to the single-photon level and SMF coupled. The repetition rate is $80$ MHz, which results in single photons separated by approximately $12$ ns. There is no spectral correlation between photons from consecutive pulses. We record the timing information and post-process this data as outlined above. We then plot the histogram of the array counts and compare them with the spectrum measured by commercial spectrometer, as shown in figure \figref{fig:TIF} (a). Next, we use the same data to see the spectral correlation between two coherent states originating in attenuated consecutive pulses. This histogram is depicted in  \figref{fig:TIF} (b).  As expected, the histogram is perfectly symmetric, showing no spectral correlations between the two photons. 

We demonstrate a multi-photon correlation spectrometer based on a grating monochromator and an array of single photon avalanche diodes. It is a step forward in the accurate characterization of single-photon sources, and is useful in many areas of quantum optics. This technique can be used in combination with existing commercial spectrometers for increased accuracy in measurement of single-photon spectra or multi-photon correlation measurement.

The authors  acknowledge  funding from NSERC (Discovery, USRA, CGS), Ontario Ministry of Research and Innovation (ERA program), CIFAR, Industry Canada and the CFI. The research leading to these results has received funding from the European Union Seventh Framework Programme (FP7/2007-2013) under grant agreement no.~257646. The authors thank Simone Tisa and Franco Zappa for their contribution in developing the detectors array, Christopher Erven for lending the SPDC source and Rolf Horn for insightful discussions. PK acknowledges support by Foundation for Polish Science under Homing Plus no.~2013-7/9 program supported by European Union under PO IG project, and by NCU internal grant no.~\mbox{1625-F}, and by the National Laboratory FAMO in Torun, Poland.


%
%
%
%
%


%

\end{document}